\documentstyle[amssymb,aps,preprint,epsf]{revtex}

\tightenlines

\begin{document}
\title{Coalescence in the 1D Cahn-Hilliard model\\
Processus de coalescence pour l'\'{e}quation de Cahn-Hilliard
unidimensionnelle}
\author{Simon Villain-Guillot}
\address{Centre de Physique Mol\'{e}culaire Optique et Hertzienne,\\
Universit\'{e} Bordeaux I, 33406 Talence Cedex, France\\
s.villain@cpmoh.u-bordeaux1.fr}
\maketitle

\begin{abstract}
We present an approximate analytical solution of the Cahn-Hilliard equation
describing the coalescence during a first order phase transition. We have
identified all the intermediate profiles, stationary solutions of the
noiseless Cahn-Hilliard equation. Using properties of the soliton lattices,
periodic solutions of the Ginzburg-Landau equation, we have construct a
family of ansatz describing continuously the process of destabilization and
period doubling predicted in Langer's self similar scenario\cite{langer}.
\end{abstract}

Pacs numbers~: 05.45.Yv, 47.20.Ky, 47.54.+r

\section{Introduction}

When a homogenous system departs suddenly from equilibrium, the fluctuations
around the initial ground state are linearly amplified and the homogenous
phase can for example spontaneously separate into two different more stable
states. The interfaces which delimit the numerous resulting monophasic
domains will interact with each other, either giving rise to formation of a
complex pattern, or merging into a single interface when the domains of the
same state slowly coalesce, minimizing the overall interfacial energy. It
results then in only two well separated domains. This process of first order
phase transition arises particularly for binary mixtures\cite{wagner} alloys 
\cite{hillert}, or vapor condensation \cite{beg}.

It can either initiate via a nucleation process, where the homogenous state
is put suddenly in a metastable configuration, and an energy barrier has to
be crossed before the transition appears. Or via a spinodal decomposition
when the system is led into a linearly unstable configuration. In this
latter case, the leading instability selects a modulation of the order
parameter at a well defined length scale, which will grows and, due to
non-linearities, saturates. The resulting pattern is composed of well
defined interfaces delimiting domains containing one of the two stable
phases. Remarkably, this intermediate stage conserves the modulation width,
and the resulting stationary pattern is of almost the same length scale as
the one selected initially \cite{izu,pre}. The dynamics finally ends with a
much slower, self-inhibiting process, dominated by the interactions between
the interfaces. The different regions of each phase coalesce in the
so-called Ostwald ripening where the number of domains diminishes whereas
their typical size increases. The asymptotic state is decomposed into two
domains, one for each phase. In this article, by spinodal decomposition, we
refer to the first stage of the dynamics only, while coarsening will denote
the second stage. Although this coarsening dynamics is in fact already
present, its influence can be often neglected during the first stage of the
process.\newline

Hillert\cite{hillert}, Cahn and Hilliard\cite{CH} have proposed a model
equation describing the segregation for a binary mixture. This model, known
as the Cahn-Hilliard equation (C-H later on), belongs to the Model B class
in Hohenberg and Halperin's classification \cite{halperin}. It is a standard
model for phase transition with conserved quantities and has applications to
phase transition in liquid crystals\cite{coullet}, segregation of granular
mixtures in a rotating drum\cite{oyama} , or formation of sand ripples \cite
{melo,stegner}. It is a partial differential equation to which a
conservative noise is added to account for thermal fluctuations\cite{cook}.%
\newline

Figure (\ref{fig1}) shows snapshots of a numerical integration of the (C-H)
dynamics which represents the full phase transition process after a quench
in temperature. Thermal fluctuations were present in the initial conditions,
but have been omitted in the dynamics. The three main stages of the spinodal
decomposition described above are clearly distinguished: first, from Fig. (%
\ref{fig1} (a)) to Fig. (\ref{fig1} (b)), we observe the selection of a
typical length scale for the modulations, then, from Fig. (\ref{fig1} (b))
to Fig. (\ref{fig1} (c)), the non-linear growth and its saturation. We note
that the number of peaks has been almost conserved between these two
configurations. On the contrary, during the coarsening dynamics observed
between Fig. (\ref{fig1} (c)) and Fig. (\ref{fig1} (d)) the typical length
of the pattern is increasing while on the other hand the amplitude of the
modulation slowly growth to reach its asymptotic value.

An important activity has been devoted to the description of the dynamics of
phase transition, using both statistical methods or numerical simulations
(for a review see \cite{bray}). The late stage of the spinodal decomposition
where the coarsening dynamics dominates exhibits ''dynamical scaling''~: the
dynamics presents a self-similar evolution where time enters only through a
length scale $L(t)$, associated with a typical length of the domains or the
rate of decay of the inhomogeneities. For instance, scaling arguments and
stability criteria give the law $L(t)\sim t^{1/3}$ for spatial dimensions
greater than one and a logarithmic behavior for one dimension in the case of
the (C-H) equation \cite{bray}.

This last stage, as observed in two-dimensional demixion of copolymer\cite
{copol} and as suggested initially by Langer\cite{langer}, can be described
as a self similar process of synchronous fusion and evaporation of domains.
This observation motivated our work and the aim of this article is to
present a one dimensional ansatz describing continuously the coalescence
process. This ansatz is in the form of a one parameter family of symmetric
profiles which interpolates between two stationary states composed of
homogeneous domains of length $\lambda /2$ and $\lambda $. It allows to
realize a self similar sequence of coalescence process in 1D, starting from
the regular micro phase separated states issued from the non-linear
saturation of the spinodal decomposition dynamics and ending with the single
interface which characterize the infinite time, thermodynamic stable state.

The paper is organized as followed: 
first, we present a brief review on general properties of phase segregations
and on the (C-H) model, mainly to fix the notations. We will reproduce
briefly the original derivation by Cahn and Hilliard, restricting ourselves
to the one dimensional case. 
In part III, we present a family of symmetric solutions of the
Ginzburg-Landau equation which is used to study the dynamics of spinodal
decomposition and to determine all the symmetric stationary state of the
(C-H) dynamics. Then in part IV, the main original part of this work, we
introduce a non-symmetric family of solutions of the (G-L) equation which is
used to construct a continuous interpolation between two consecutive
symmetric stationary states. After a study of the energy landscape
associated with this ansatz, we finally discuss the numerical accuracy of
our calculations. In the conclusion, we justify the hypothesis we have made
and compare the suggested scenario with coalescence in real systems.

\section{The Cahn-Hilliard model}

The Cahn-Hilliard theory is a modified diffusion equation; it is a
continuous conservative model for the scalar order parameter $\Phi $, which
reads in its dimensionless form:

\begin{equation}
\frac{\partial \Phi }{\partial t}\left( {\bf r},t\right) ={\bf \nabla }^{2}(%
\frac{\varepsilon }{2}\Phi +2\Phi ^{3}-{\bf \nabla }^{2}\Phi )+\xi \left( 
{\bf r},t\right) .  \label{CHeq}
\end{equation}
The real order parameter can correspond to the dimensionless magnetization
in Ising ferromagnet, to the fluctuation of density of a fluid around its
mean value during a phase separation or to the concentration in some region
around ${\bf r}$ of one of the components of a binary solution. $\varepsilon 
$ is the dimensionless control parameter of the system ; it is often
identified to the reduced temperature ($\varepsilon =\frac{T-T_{c}}{T_{c}}$
where $T_{c}$ is the critical temperature of the phase transition). This
equation, first derived by Cahn and Hilliard \cite{CH}, has also been
retrieved by Langer\cite{langer} from microscopic considerations. As
written, the (C-H) equation does account for thermal fluctuations present in
the system through a random white noise $\xi \left( {\bf r},t\right) $,
whose amplitude is proportional to the square root of the temperature of the
system.\newline

The homogeneous stationary solutions for the noiseless (C-H) equation are
extrema of the effective Ginzburg-Landau potential $V(\Phi )=\frac{%
\varepsilon }{2}\Phi ^{2}+\Phi ^{4}$ (G-L later on). For positive $%
\varepsilon $, there is only one homogenous solution $\Phi =0$ which is
linearly stable; for negative $\varepsilon $, the stationary solution $\Phi
=0$ undergoes a pitchfork bifurcation and three stationary solutions exist. $%
\Phi =0$ is still a stationary solution, but it is now linearly unstable ;
two other symmetric solutions $\Phi =\pm \frac{\sqrt{-\varepsilon }}{2}$ are
stable and have the same free energy $F=-\varepsilon ^{2}/32$. Thus, a first
order transition can be experienced by quenching the system suddenly from a
positive reduced temperature $\varepsilon $ to a negative one. Spinodal
decomposition is the resulting dynamics.

The stability of the solution $\Phi =0$ can be studied by linearizing
equation (\ref{CHeq}) around $\Phi =0$ (i.e. neglecting the non linear term $%
\Phi ^{3}$); considering $\Phi $ as a sum of Fourier modes:

\begin{equation}
\Phi ({\bf r},t)=\sum_{{\bf q}}\phi _{{\bf q}}e^{i{\bf q\cdot r}+\sigma t}
\end{equation}
where $\phi _{q}$ is the Fourier coefficient at $t=0$, we obtain for the
amplification factor $\sigma (q)$ :

\begin{equation}
\sigma ({\bf q})=-(q^{2}+\frac{\varepsilon }{2})q^{2}
\end{equation}
It shows immediately that $\Phi =0$ is linearly stable for $\varepsilon >0$
while a band of Fourier modes are unstable for negative $\varepsilon $,
since $\sigma ({\bf q})>0$ for $0<q<\sqrt{(-\varepsilon /2)}$. Moreover, the
most unstable mode is for $q_{C-H}=\sqrt{-\varepsilon }/2$(with $\sigma
_{\max }=\frac{\varepsilon ^{2}}{16}$). This wave number of maximum
amplification factor will dominate the first stage of the dynamics; in
particular, it explains why the modulations appear at length scales close to 
$\lambda _{C-H}=2\pi /q_{C-H}$, the associated wave length. Later on,
interfaces separating each domain interact through coalescence dynamics,
causing $<\lambda >$ to change slowly toward higher values \cite{baron,izu}

We will now use known results on non-homogeneous solutions of the (G-L)
equation to study both the saturation of the spinodal decomposition and the
coalescence.

\section{Stationary States of the Cahn-Hilliard Dynamics}

\subsection{\protect\bigskip Symmetric Soliton Lattice Solutions}

For $\varepsilon <0$, \ there exists a stationary solution of the one
dimensional (C-H) that relies the two homogenous phases $\Phi =\pm \frac{%
\sqrt{-\varepsilon }}{2}$

\begin{equation}
\Phi (x)=\frac{\sqrt{\left| \varepsilon \right| }}{2}\tanh (\frac{\sqrt{%
\left| \varepsilon \right| }x}{2}).  \label{tanh}
\end{equation}
Such a monotonic solution describes a continuum interface between the two
stable homogeneous phases, and corresponds to the thermodynamically stable
solution that ends the phase transition dynamics. But this is a particular
member of a one parameter family of stationary solutions of the (G-L)
equation

\begin{equation}
\frac{\varepsilon }{2}\Psi +2\Psi ^{3}-\nabla ^{2}\Psi =0  \label{CH0}
\end{equation}
These solutions, the so-called soliton-lattice solutions, are~: 
\begin{equation}
\Psi _{k,\varepsilon }(x)=k\Delta {\rm Sn}(\frac{x}{\xi },k)\text{ with }\xi
=\Delta ^{-1}=\sqrt{2\frac{k^{2}+1}{-\varepsilon }}  \label{amplitude}
\end{equation}
where ${\rm Sn}(x,k)$ is the Jacobian elliptic function sine-amplitude, or
cnoidal mode. This family of solutions is parametrized by $\varepsilon
^{\ast }$ and by the modulus $k\in \left[ 0,1\right] $, or ''segregation
parameter''. These solutions describe periodic patterns of periods

\begin{equation}
\lambda =4K(k)\xi \text{, where }K(k)=\int_{0}^{\frac{\pi }{2}}\frac{{\rm d}t%
}{\sqrt{1-k^{2}\sin ^{2}t}}  \label{period}
\end{equation}
is the complete Jacobian elliptic integral of the first kind. Together with $%
k$, it characterizes the segregation, defined as the ratio between the size
of the homogeneous domains, $0.5\times \lambda $, and the width of the
interface separating them, $2\times \xi $. The equation (\ref{period}) and
the\ relation $\xi =\Delta ^{-1}$, enable to rewrite this family as : 
\begin{equation}
\Psi _{k,\lambda }(x)=\frac{4K(k)\cdot k}{\lambda }{\rm Sn}(\frac{4K(k)}{%
\lambda }x,k).
\end{equation}
This family of profiles (or alternating interfaces) can be obtain exactly as
a periodic sum of single solitons and antisolitons\cite{saxena} 
\begin{equation}
\sum_{n}(-1)^{n}\tanh (\pi s(x-n))=\frac{2k(s)K(s)}{\pi s}{\rm Sn}(x,k)%
\hbox{ with  }s=\frac{K(k)}{K(k^{\prime })}\text{ and }k^{\prime 2}=1-k^{2}
\label{sum}
\end{equation}

\subsection{Antsatz for the Spinodal Decomposition Dynamics\qquad}

The preceding family of profiles can be used to explore the spinodal
decomposition dynamics. It can be associated with a micro phase separation,
locally limited by the finite diffusion coefficient. For $k=1$, ${\rm Sn}%
(x,1)=\tanh (x)$, we recover the usual single interface solution~(\ref{tanh}%
), of width $2/\sqrt{\left| \varepsilon \right| }$; it is associated with a
one soliton solution and corresponds to a strong, or macroscopic
segregation. Note that $K(1)$ diverges~; the solution 
\begin{equation}
\Psi _{1,\varepsilon }(x)=\frac{\sqrt{\left| \varepsilon \right| }}{2}\tanh (%
\frac{\sqrt{\left| \varepsilon \right| }}{2}x).
\end{equation}
is thus the limit of infinite $s$, when the solitons, entering in relation (%
\ref{sum}), are far apart one each others. In the opposite limit (weak
segregation regime), it describes a sinusoidal modulation 
\begin{equation}
{\rm lim}_{k\rightarrow 0}\Psi _{k,\varepsilon }(x)=k\sqrt{\frac{\left|
\varepsilon \right| }{2}}\sin (\sqrt{\frac{\left| \varepsilon \right| }{2}}%
x)=k\frac{2\pi }{\lambda }\sin (\frac{2\pi }{\lambda }x)=kq\sin (qx)
\end{equation}
It will correspond to the Fourier mode $q=\frac{2\pi }{\lambda }$ of the
initial white noise, with an arbitrary small amplitude $\nu =kq$. Since
experiences, numerical simulations and linear stability analysis show that $%
\lambda $, the spatial period of the pattern is constant during the whole
spinodal decomposition process, we choose $\lambda $ to coincide with the
most instable wave length obtain with the Cahn Hilliard linear approach, $%
\lambda =\lambda _{C-H}=\frac{4\pi }{\sqrt{-\varepsilon _{0}}}$, where $%
\varepsilon _{0}$ is the quench temperature. Thus, we obtain a one parameter
family of profiles $\Psi ^{\ast }(x,k)=\Psi _{k,\lambda _{C-H}}(x)$ which
describe very well both the linear growth and the saturation . The dynamics
is now reduced to the time evolution of the single free parameter : $k(t)$.\
Using equations (\ref{amplitude}) and (\ref{period}), we find that $\lambda $%
, $k$ and $\varepsilon $ are related to one another through the state
equation 
\begin{equation}
\varepsilon (k)=-2(1+k^{2})\left( \frac{4K(k)}{\lambda }\right) ^{2}.
\label{implicit}
\end{equation}
So, this implicit equation tells us that if we fixe $\lambda =\lambda _{C-H}$%
, the dynamics can also be reduced to the evolution of $\varepsilon (t)$.

Given a periodic function $\Phi $ (obtained either from experimental data or
numerical simulation of equation (\ref{CH0})) at time $t$, the {\it ansatz}
assumes that it corresponds to a soliton lattice of the same period: i.e.,\
there exists $k(t)$ such that $\Psi _{k,\lambda _{C-H}}(x)\sim \Phi (x,t)$
for each time $t$. For this purpose, we have developed three different
algorithms, taking advantage of the general properties of the family of
solutions $\Psi ^{\ast }(x,k)$ ~: either, $k$ can be deduced both from the
amplitude of the oscillation equals to $4kK(k)/{\lambda }$, or from the
relation $k=\!1-\left( (\Psi ({\lambda }/2,k)/\Psi ({\lambda }%
/4,k))^{2}-1\right) ^{2}$; thirdly, a straightforward computation relates $k$
to the ratio of the two first Fourier modes of the soliton lattice $\Psi
^{\ast }({x},k)$. We have observed that the three methods show in general
similar results within an error of one percent.

In this approach, $\varepsilon (t)$ can be then interpreted as a fictitious
temperature or ``local temperature'' of the domains: it is the temperature
extracted from the profile at a given time, using the correspondence between 
$\varepsilon $ and $k$ of equation (\ref{implicit}). For instance, at $t=0$,
the amplitude is small and we find that $k(0)=\frac{\nu \lambda _{m}}{2\pi }%
\rightarrow 0$ and thus $\varepsilon ^{\ast }(0)=8\pi ^{2}/{\lambda }^{2}$,
different {\it a priori} from $\varepsilon _{0}$ ($\varepsilon ^{\ast }(0)=%
\frac{\varepsilon _{0}}{2}$ for $\lambda =\lambda _{C-H}$).

Somehow, the dynamics of (C-H) can be projected at first order onto a
dynamics along the sub-family $\Psi ^{\ast }(x,k)=\Psi _{k,\lambda
_{C-H}}(x) $, which can be considered as an attractor of the solutions, i.e.
the density profile of the system will evolve with time, staying always
close to a function $\Psi ^{\ast }(x,k)$. And using a solubility condition,
it is possible to compute the full non linear part of this dynamics, the
saturation of the spinodal decomposition, which leads the system in a well
defined stationary state\cite{pre}.

\subsection{Saturations of the Spinodal Decomposition Dynamics}

According to the previous interpretation of the parameter $\varepsilon $, as 
$\varepsilon (t=0)=\frac{\varepsilon _{0}}{2}$, the system is initially out
of equilibrium. The dynamics will saturate when this fictitious temperature
will reach the real thermodynamic one, i.e. the quench temperature $%
\varepsilon _{0}$; that is, using equation of state (\ref{implicit}) for $%
\lambda =\lambda _{C-H}$, when $k=k^{s}$ solution of the implicit equation : 
\begin{equation}
2(1+k^{s2})K(k^{s})^{2}=-\frac{\varepsilon _{0}\lambda _{{\small C-H}}^{2}}{%
16}=\pi ^{2}\text{ \quad that is\quad\ }k^{s}\!=\!0.687
\end{equation}
Note that in this case, the width of the interface, which was initially,
just after the quench, proportional to $\frac{2}{\sqrt{-\varepsilon _{0}}}$
has now become proportional to $\frac{\pi }{\sqrt{-\varepsilon _{0}}K(k^{s})}%
=\frac{\sqrt{2(1+k^{s}\!^{2})}}{\sqrt{-\varepsilon _{0}}}\backsimeq \frac{1.7%
}{\sqrt{-\varepsilon _{0}}}$: the segregation has slightly increased. Using
linear stability analysis, Langer has shown that the profile thus obtained, $%
\Psi ^{\ast }(x,k^{s})=\Psi (x,k^{s},\lambda _{C-H})$, is destroyed by
stochastic thermal fluctuations and he has identified the most instable mode
as an ''antiferro'' mode, leading to a period doubling. The result of this
destabilization is another profile of alternate interface, where the length
of the domains is now :$\lambda =2\lambda _{C-H}=\frac{8\pi }{\sqrt{%
-\varepsilon _{0}}}$. This means that the new stationary profile is given by 
$\Psi (x,k_{1}^{s},2\lambda _{C-H})$, where $k_{1}^{s}$ is solution of the
implicit relation

\begin{equation}
2(1+k_{1}^{s2})K(k_{1}^{s})^{2}=-\frac{\varepsilon _{0}(2\lambda _{{\small %
C-H}})^{2}}{16}=4\pi ^{2}=8(1+k^{s2})K(k^{s})^{2}\text{ \quad that is\quad\ }%
k_{1}^{s}\!=\!0.985
\end{equation}
The interface of this new profile is relatively sharper (the width of the
interface is now proportional to $\frac{2\pi }{\sqrt{-\varepsilon _{0}}%
K(k_{2}^{s})}=\frac{\sqrt{1+k_{1}^{s}\!^{2}}}{\sqrt{2}\sqrt{-\varepsilon _{0}%
}}\backsimeq \frac{2\times 0.99}{\sqrt{-\varepsilon _{0}}}$) compare to the
size of the homogeneous domains which has now double, see Fig. (\ref{fig2}).%
%
%
%
%
%
%
%
Again, this new stationary profile turns out to be linearly instable with
respect to an ''antiferro'' perturbation of period $4\lambda _{C-H}$.

Thus these families of profiles and instabilities enable to describe the one
dimensional coarsening as a cascade of doubling process, leading from a
pattern of wave length $\lambda _{C-H}$ composed of domains separated by
interfaces to a single ${\rm tanh}(\frac{\sqrt{-\varepsilon _{0}}}{2}x)$
interface separating two semi infinite domains. Each of these successive
intermediate profiles can be described by an element of the above family of
soliton lattice $\Psi (x,k_{n}^{s},2^{n}\times \lambda _{C-H})$. We thus
have a family of segregation parameter $\left\{ k_{n}^{s}\right\} $, which
are determined by the implicit relations 
\begin{equation}
2(1+k_{n}^{s2})K(k_{n}^{s})^{2}=-\frac{\varepsilon _{0}(2^{n}\lambda _{%
{\small C-H}})^{2}}{16}=\pi ^{2}2^{2n}.  \label{implicitn}
\end{equation}
We have found numerically for the first of them 
\begin{equation}
\begin{tabular}{|l|}
\hline
$k_{0}^{s}\!=k^{s}\!=\!0.6869795924$ \\ \hline
$k_{1}^{s}\!=\!0.9851675587$ \\ \hline
$k_{2}^{s}\!=0.99997210165$ \\ \hline
$k_{3}^{s}\!=\!0.9999999999027$ \\ \hline
\end{tabular}
\begin{tabular}{|l|}
\hline
$k_{0}^{s}\Delta _{0}^{s}\!=\!0.400\sqrt{-\varepsilon _{0}}$ \\ \hline
$k_{1}^{s}\Delta _{1}^{s}\!=\!0.496250\sqrt{-\varepsilon _{0}}$ \\ \hline
$k_{2}^{s}\Delta _{2}^{s}\!=0.499990\sqrt{-\varepsilon _{0}}$ \\ \hline
$k_{3}^{s}\Delta _{3}^{s}\!=\!0.49999846\sqrt{-\varepsilon _{0}}$ \\ \hline
\end{tabular}
\end{equation}
We see that $\left\{ k_{n}^{s}\right\} $ converges toward $k_{\infty }^{s}=1$
(single interface case) ; meanwhile the amplitude of the modulation $%
k_{n}^{s}\Delta _{n}^{s}$ goes toward $\sqrt{\left| \varepsilon _{0}\right| }%
/2$, as can be seen in the second column of the table above. For large $n$,
we can conclude from the implicit relation (\ref{implicitn}) that the ratio
of the domain size to the interface width characterized by $K(k_{n}^{s})$
behaves as $\pi 2^{n-1}$. Each of the stationary profiles 
\begin{equation}
\Psi _{n}(x)=\Psi (x,k_{n}^{s},2^{n}\lambda _{{\small C-H}})=\frac{\sqrt{%
-\varepsilon _{0}}K(k_{n}^{s})\cdot k_{n}^{s}}{2^{n}\pi }{\rm Sn}(\frac{%
\sqrt{-\varepsilon _{0}}K(k_{n}^{s})}{2^{n}\pi }x,k_{n}^{s}),
\end{equation}
for which the interface width is proportional to $\frac{2^{n}\pi }{\sqrt{%
-\varepsilon _{0}}K(k_{n}^{s})}$(which tends to $\frac{2}{\sqrt{-\varepsilon
_{0}}}$, in agreement with ${\rm tanh}(\frac{\sqrt{-\varepsilon _{0}}}{2}x)$%
), is identically destroyed by the Langer ''antiferro'' instability.

\section{An Ansatz for the 1D coarsening process}

\subsection{Non-symmetric soliton lattice Profile}

In order to describe one step of the coalescence process, i.e. the dynamics
that start from $\Psi _{n}(x)$ and ends with the profile $\Psi _{n+1}(x)$ ,
we will use another family of equilibrium profiles\cite{novik}, solutions of
(G-L) equation, which write: 
\begin{equation}
\psi (a,k,x)=\frac{\alpha (a,k)-k/\sqrt{a}\beta (a,k)Sn(4x\frac{K(k)}{%
\lambda },k)}{1-k/\sqrt{a}Sn(4x\frac{K(k)}{\lambda },k)}
\end{equation}
where $\alpha (a,k)=\frac{-2k^{2}/a+1+k^{2}}{%
((1+k^{2})^{2}-12k^{2}+2(a+k^{2}/a)(1+k^{2}))^{\frac{1}{2}}}$ \ and $\ \beta
(a,k)=\frac{2a-1-k^{2}}{((1+k^{2})^{2}-12k^{2}+2(a+k^{2}/a)(1+k^{2}))^{\frac{%
1}{2}}}$ .

It is still a periodic lattice of interfaces, but now, the mean value of the
order parameter is non zero (non symmetric case). It is controlled by the
parameter $a\geqslant 1$ : if $a$ goes infinity, we recover the previous
family of periodic profiles.

\subsection{\protect\bigskip Ansatz for the continuous interpolation between
two stationary states}

If we choose $a$ to be equal to $1+k^{\prime }$ (where $k^{\prime 2}=1-k^{2}$%
), we can then construct symmetric profiles using the sum of two
non-symmetric ones. %
Indeed, using Gauss' transformation (or descending Landen transformation 
\cite{abra}), which relates the soliton lattice of spatial period $2\lambda $
(and of modulus $k$) to the soliton lattice of period $\lambda $ (and of
modulus $\mu =\frac{1-k^{\prime }}{1+k^{\prime }}$), we have 
\begin{eqnarray}
&& 
\begin{array}{ccc}
1-\frac{\sqrt{5-k^{2}}}{2}(\psi (k,x-\frac{\lambda }{2})+\psi (k,x+\frac{%
\lambda }{2})) & = & kSn(2x\frac{K(k)}{\lambda },k)
\end{array}
\label{final} \\
&& 
\begin{array}{ccc}
1-\frac{\sqrt{5-k^{2}}}{2}(\psi (k,x-\lambda )+\psi (k,x+\lambda )) & = & 
(1-k^{\prime })Sn((4x+2\lambda )\frac{K(\mu )}{\lambda },\mu )
\end{array}
\label{initial}
\end{eqnarray}
where $\psi (k,x)=\psi (1+k^{^{\prime }},k,x)$. Thus, we then can show from
equation (\ref{final}) that 
\begin{equation}
K(k)\left[ 1-\frac{\sqrt{5-k^{2}}}{2}(\psi (k,x-\frac{\lambda }{2})+\psi
(k,x+\frac{\lambda }{2}))\right] =kK(k)Sn(2x\frac{K(k)}{\lambda },k).
\end{equation}
This is the solution of the G-L equation of period $2\lambda $. Moreover, if
we use the fact that 
\begin{equation}
K(k)=\frac{2}{1+k^{\prime }}K(\mu )\text{ or }K(\mu )=\frac{1}{1+\mu }K(k)%
\text{ }  \label{exacte}
\end{equation}
we can write 
\begin{equation}
(1-k^{\prime })K(k)Sn((4x+2\lambda )\frac{K(\mu )}{\lambda },\mu )=2\mu
K(\mu )Sn((2x+\lambda )\frac{2K(\mu )}{\lambda },\mu ).
\end{equation}
Then, using relation (\ref{initial}), the solution of the (G-L) equation of
period $\lambda $ can be expressed as 
\begin{equation}
K(k)\left[ 1-\frac{\sqrt{5-k^{2}}}{2}(\psi (k,x-\lambda )+\psi (k,x+\lambda
))\right] =2\mu K(\mu )Sn((2x+\lambda )\frac{2K(\mu )}{\lambda },\mu ).
\end{equation}
So, we see that both the initial state $\Psi ^{\ast
}(x,k_{n-1}^{s},2^{n-1}\lambda _{C-H})$ and the final state $\Psi ^{\ast
}(x,k_{n}^{s},2^{n}\lambda _{C-H})$ of a step of the coalescence process can
be describe, modulo a phase shift, by the same function : 
\begin{equation}
\Phi (x,k,\phi )=\frac{4K(k)}{\lambda }\left[ 1-\frac{\sqrt{5-k^{2}}}{2}%
(\psi (k,x-(1-\phi /2)\lambda )+\psi (k,x+(1-\phi /2)\lambda ))\right]
\end{equation}
with $k=k_{n}^{s}$ and $\lambda =2^{n}\lambda _{C-H}$. Therefore we can
describe the coalescence by a transformation at constant segregation
parameter $k$, while the degree of freedom $\phi $, associated with the
relative phase between the two profiles, evolves in time from $0$ to $1$
according to the C-H dynamics.

This non-symmetric lattice of interfaces can be interpreted as a periodic
sum of alternating single interfaces (kinks and antikinks). In the same
spirit as relation (\ref{sum}), if one forget in the infinite sum every two
out of four interfaces, one gets : 
\begin{equation}
\psi (x)\backsim \sum_{p}\left[ \tanh (\pi s(x-4\ast p))-\tanh (\pi
s(x-4\ast p+1))\right] .
\end{equation}
Then (see Fig.(\ref{fig3}))\ adding $\psi (x+2)$ to $\psi (x)$ enables to
recover relation (\ref{sum}), while, after a translation, adding $\psi (x+1)$
and $\psi (x)$ gives the soliton lattice of double period, because of the
cancelation of half of the interfaces (annihilation of kinks and antikinks).

If we look at the time evolution of the profile $\Phi (x,k,\phi )$, starting
from the region $\phi =0$, we can transform the (C-H) equation into a phase
field equation, replacing $\frac{\partial }{\partial t}$ $\Phi (x,k,\phi )$
\ by $\frac{\partial }{\partial \phi }\Phi (x,k,\phi (t)).\frac{d\phi }{dt}$%
. The dynamics will be similar to a spinodal decomposition, with $\phi $
growing and saturating exponentially. $\frac{\partial }{\partial \phi }\Phi
(x,k,\phi )$ is the most unstable mode founded in Langer's linear stability
analysis, characterized by the alternate growth and decrease of domains
(''antiferro'' mode). Note that when Langer was studying the most instable
perturbation, he was looking at the linearized version of C-H equation
around $\Psi ^{\ast }(k,x)=\Psi (x,k,\lambda _{C-H})$ : 
\begin{equation}
{\cal L}\varphi =\left( \frac{\varepsilon }{2}+6\Psi ^{\ast 2}-\nabla
^{2}\right) \varphi =\left( \frac{\varepsilon }{2}+n\times (n+1)\Psi ^{\ast
2}-\nabla ^{2}\right) \varphi .
\end{equation}
${\cal L}\varphi =E\varphi $ is the Lam\'{e} equation, for $n=2$ (here $%
\varepsilon _{0}=1)$. This equation doesn't have simple (algebraic) exact
eigenfunction of period $2\lambda _{C-H}$ \cite{arscott}. $\frac{\partial }{%
\partial \phi }$ $\Phi (x,k,\phi )$ for $\phi =0$ is not an exact
eigenfunction either \cite{annexe}. Nevertheless, it happens to be a good
approximation for the eigenfunction of lowest eigenvalue. Due to the
concavity of ${\cal F}(\phi )$ around $\phi =0$ (see below Figure (\ref{fig5}%
)), this eingenvalue will be negative, triggering a linear destabilization
and an exponential amplification of the perturbation, i.e. an exponential
growth of the translation $\phi $ with time.

Langer's phenomenon of ''antiferro'' instability appears due to the
existence of two possible directions for displacement of the interfaces $%
"\tanh "$ (or of the non-symmetric lattice of interfaces $\psi $), one with
a positive velocity ($+\frac{d\phi }{dt}$) and one with a negative one ($-%
\frac{d\phi }{dt}$). The four different kinds of interfaces present in a
cell of length $2\lambda _{C-H}$ have alternately a positive or a negative
velocity. This can be seen as the existence of two antisymmetric patterns 
\cite{goldstein}, or building blocks for the leading instability around a
intermediate state $\Psi ^{\ast }(x,k_{n}^{s},2^{n-1}\times \lambda _{C-H})$
(see Figure (\ref{fig4})). $\pm \frac{d}{dx}\psi (x)$, these two building
blocks, are associated with the two pairs of interfaces, $\psi (x)$ and $%
\psi (x+2)$ which have been used to construct our ansatz.

Note that in Langer's analysis, the breaking of symmetry for the choice of
the antiferro cell, corresponds here to the freedom we have when choosing
the range of variation of $\phi $ : we could have chosen to go from $0$ to $%
-1$, ending after a step of coarsening with the symmetric pattern, or
equivalently, a pattern translated of half a period.

\subsection{Energy Landscape}

In order to prove the usefulness of this ansatz, we have plot the energy
averaged over the final period, ${\cal F}(\phi )=\int F(\Phi (x,k,\phi ))%
{\rm d}x$, as a function of the parameter $\phi $, keeping $k$ constant. We
see for example in Figure (\ref{fig5}) that the value $\phi =0$ correspond
to a local maximum of energy, while $\phi =1$ (or $-1$) is a minimum. Note
that there is no energy barrier in this particular energy landscape, in
agreement with linear stability analysis.%

\subsection{Approximation for adiabatic evolution with constant $k$}

Relation (\ref{implicitn}), enables to write an implicit relation between $%
k_{n}$ and $k_{n+1}$ : 
\begin{equation}
\sqrt{1+k_{n+1}^{s2}}K(k_{n+1}^{s})=2\sqrt{1+k_{n}^{s2}}K(k_{n}^{s})
\label{implicitK}
\end{equation}
It is in fact different from the relation (\ref{exacte})\ obtained by the
Landen transformation. Nevertheless in the region $k\geq k_{0}^{s}\!=\!0.687$%
, the two transformation almost coincide, as can be seen on Figure (\ref
{fig6}).\ As this is especially true close to $k=1$, and as this region is
rapidly reach after the second or third iteration, we pretend that the
process of coarsening can be describe with a reasonable accuracy by our
antsatz at constant $k$. A slight change of segregation parameters during
the $n^{th}$ doubling process, from $k=Landen(k_{n}^{s})$ when $\phi =0$ to $%
k=k_{n+1}^{s}$ when $\phi =1$, is present in the dynamics, as seen in the
following table 
\begin{equation}
\begin{tabular}{|l|l|}
\hline
$k_{0}^{s}\!=\!0.6869795924$ & $Landen^{-1}(k_{1}^{s}%
\!)=Gauss(k_{1}^{s})=0.7070743852$ \\ \hline
$Landen(k_{0}^{s}\!)=0.9826346738$ & $k_{1}^{s}\!=\!0.9851675587$ \\ \hline
$Landen(k_{1}^{s}\!)=0.9999720868$ & $k_{2}^{s}\!=0.99997210165$ \\ \hline
$Landen(k_{2}^{s}\!)=0.9999999998$ & $k_{3}^{s}\!=\!0.999999999902745$ \\ 
\hline
\end{tabular}
\end{equation}
But it has only a minor effect in terms of profile shape $\Phi (x,k,\phi )$.
This means that the dynamics of the parameter $k$, which affect the value of
the modulus by $0.25$\% during the first step of coarsening, becomes
negligeable as $k$ goes closer to $1$. So we can conclude that this ''$k$%
-dynamics'' is irrelevant ; this parameter can be considered as constant
during the evolution of $\phi $.%

\section{Discussion on the hypothesis}

Our analytic method rely on the assumption that at each step of the
dynamics, the system can be characterized by a specific spatial period : we
need therefore to discuss how this approach is relevant to the general case
where noise is present. We have noted numerically that, for the spinodal
decomposition, the average size of the modulation is $\lambda _{C-H}$, with
a deviation of less than one percent from the value predicted by the linear
theory. It does not mean that, in a real system, each domain has a length
scale of $\lambda _{C-H}$, but that the distribution of the domains' length
will be centered around $\lambda _{C-H}$ \cite{pre}. The coalescence events,
due to initial fluctuations in the periodicity of the pattern selected just
after the quench, can be neglected during the initial growth of the
amplitude of the modulation. Only after this initial growth has saturated
(as can be seen from Figure (\ref{fig1})), starts the coalescence process,
which will then dominate the dynamics : the typical length scale of the
structures increases slowly with time.

As suggested by \cite{langer,copol}, we can suppose that during the ideal
coalescence process, each lattice of interfaces will experience an antiferro
instability. By ideal coalescence, we mean a process which breaks as few
symmetries as possible. But in a real system, this instability will concern
a region of finite size, where it choose a certain sublattice, or a range
for $\phi $ (for example, $\phi $ varies from $0$ to $1$), while it is the
opposite choice in the neighboring region ( $\phi $ varies from 0 to -$1$).
Thus on the overall, the global symmetry is recover. During each step of the
process, the width of the domains will locally double. But due to non
synchronization between regions, for the system as a whole, the average
length scale will vary continuously.%

These results indebted to Christophe Josserand for many fruitful discussions
and comments together with numerical help.

\begin{figure}[h]
\caption{{\protect\small Time evolution of the order parameter $\Phi (x,t)$
for $\protect\varepsilon =-1$, $dx=0.1227$. (a) initial conditions at $t=0$
are taken randomly with a very low amplitude ($5\cdot 10^{-4}$)~; (b) at
time $t=15$, the amplitude of the small scale spatial modulations has been
damped by the (C-H) dynamics, while only long wavelength contributions are
still present. ; (c) at $t=225$, the spatial modulation has almost reached
its final amplitude, keeping roughly the same number of peaks as before~;
(d) at $t=1800$, we observe that the number of domains has decreased from
the coarsening dynamics.}}
\label{fig1}
\end{figure}

\begin{figure}[h]
\caption{{\protect\small Profiles of the two first metastable solutions of
the (C-H) dynamics, with }$k_{1}^{s}\!=\!0.687${\protect\small \ and }$%
k_{2}^{s}\!=\!0.985${\protect\small ,} {\protect\small corresponding to the
first coarsening process.}}
\label{fig2}
\end{figure}

\begin{figure}[h]
\caption{{\protect\small Construction of the two first steady solutions of
the (C-H) dynamics, with }$k_{1}^{s}\!=\!0.687${\protect\small \ and }$%
k_{2}^{s}\!=\!0.985,${\protect\small \ using a superposition of the
non-symmetric profile }$\protect\psi (k,x)${\protect\small , itself
stationary solution of the (C-H) equation. By changing the phase shift
between the two profiles entering into the linear combination, one obtains
two different symmetric profiles, of periods }$\protect\lambda $%
{\protect\small \ and }$2\protect\lambda .$}
\label{fig3}
\end{figure}

\begin{figure}[h]
\caption{{\protect\small Langer's most instable perturbation mode of
destabilization of the soliton lattice is identified with }$\frac{\partial }{%
\partial \protect\phi }${\protect\small \ }$\Phi (x,k,\protect\phi )$%
{\protect\small \ at }$\protect\phi =0.${\protect\small \ It is composed of
two antisymmetric patterns, plotted in dotted (plain) line, evolving toward
right (left) at velocity +}$\frac{d\protect\phi }{dt}$ {\protect\small (-}$%
\frac{d\protect\phi }{dt}${\protect\small ), causing an ''antiferro''
instability leading to a period doubling of the pattern. They are the
spatial derivative of the initial non symmetric profile }$\protect\psi (x)$%
{\protect\small \ which has been used to construct our ansatz in Figure \ref
{fig3}.}}
\label{fig4}
\end{figure}

\begin{figure}[h]
\caption{{\protect\small Profile of the free energy landscape during a
coarsening process, F(}$\protect\phi ${\protect\small ). It starts at }$%
\protect\phi =0$ {\protect\small for a configuration characterized by the
segregation ratio} $k_{1}^{s}\!=\!0.687${\protect\small \ for which the
energy per unit length is F(}$\protect\phi =0${\protect\small )}$\simeq
-0.135${\protect\small ; one sees\ that in this region, the free energy is a
concave function of }$\protect\phi ${\protect\small \ and thus, the
associated pattern is linearly instable. The elementary step of the
coarsening process ends for }$\protect\phi =1${\protect\small \ associated
with a pattern characterized by the segregation ratio }$k_{2}^{s}\!=\!0.985$ 
{\protect\small for which the energy per unit length is  F(}$\protect\phi =1$%
{\protect\small )}$\simeq -0.45${\protect\small . In the region }$\protect%
\phi =1,${\protect\small \ the free energy is a convex function of } $%
\protect\phi ${\protect\small .}}
\label{fig5}
\end{figure}

\begin{figure}[h]
\caption{{\protect\small Comparison between the Landen transformation (upper
solid line) and the implicit relation between consecutive stationary steady
states of (C-H) equation }$(1+k_{n+1}^{2})^{\frac{1}{2}}K(k_{n+1})$=2$%
(1+k_{n}^{2})^{\frac{1}{2}}K(k_{n})$ {\protect\small (lower dash line) in
the region between }$k=k_{1}^{s}\!=\!0.68${\protect\small \ and }$k=\!1,$%
{\protect\small \ corresponding to the region of interest for the coarsening
process. The Landen transformation relates the segregation parameter of a
soliton lattice of period }$\protect\lambda ${\protect\small \ with the
segregation parameter associated with a lattice of period }$2\protect\lambda 
$.{\protect\small \ It is the generalization for the cnoidal function of the
usual relation }$\sin 2\protect\theta =2\sin \protect\theta \cos \protect%
\theta ${\protect\small . The circles mark (from left to right) }$%
k_{0}^{s},Landen^{-1}(k_{1}^{s}),k_{1}^{s}${\protect\small \ , and} $%
Landen^{-1}(k_{2}^{s}).$ {\protect\small If the doubling process associated
with the coarsening were only a phase shift of }$\protect\phi $%
{\protect\small \ between 0 and }$1${\protect\small , the two curves would
have coincided, i.e. the implicit relation (\ref{implicitK}) would be
equivalent to the Landen transformation. As it is not the case, there is a }$%
k${\protect\small \ change during the doubling process, but only of a few
percent, as can be seen on the figure above. Moreover, one sees that, as the
coarsening process takes place, }$k${\protect\small \ reaches values closer
and closer to }$1${\protect\small \ for which the disagreement becomes
negligible : the }$k${\protect\small \ change becomes smaller and smaller.}}
\label{fig6}
\end{figure}


\bigskip

\begin{references}
\bibitem{langer}  J.S.~Langer, Annals of Physics {\bf 65}, 53 (1971).

\bibitem{wagner}  C.~Wagner, Z.~Electrochem.~{\bf 65}, 581 (1961).

\bibitem{hillert}  M.~Hillert, {\it Acta Met}.~{\bf 9}, 525 (1961).

\bibitem{beg}  J.S.~Langer, in {\it Solids Far From Equilibrium}, edited by
C.~Godr\`{e}che (Cambridge University Press, Cambridge, England 1992), pp.
297-363.

\bibitem{izu}  T. Izumitani and T. Hashimoto, J. Chem. Phys., {\bf 83}, 3694
(1985).

\bibitem{pre}  S. Villain-Guillot and C. Josserand, {\it Phys. Rev.} E {\bf %
66}, 036308 (2002).

\bibitem{CH}  J.W.~Cahn and J.E.~Hilliard, {\it J.~Chem.~Phys.}~{\bf 28},
258 (1958).

\bibitem{halperin}  P.C.~Hohenberg and B.I.~Halperin, {\it Rev.~Mod.~Phys.}~%
{\bf 49}, 435 (1977). See also M.C.~Cross and P.C.~Hohenberg, {\it %
Rev.~Mod.~Phys.} {\bf 65}, 851 (1993)

\bibitem{coullet}  C.~Chevallard, M.~Clerc, P.~Coullet and J.M.~Gilli,
Eur.~Phys.~J.~E {\bf 1}, 179 (2000).

\bibitem{oyama}  Y.~Oyama, {\it Bull. Inst. Phys. Chem. Res. Rep.} {\bf 5},
600 (1939). S.~Puri and H.~Hayakawa, cond-mat/9901260 and Advances in
Complex Systems, Vol. 4, No. 4 (2001) 469-479 .

\bibitem{melo}  M.A. Scherer, F. Melo and M. Marder, Phys. Fluids {\bf 11},
58 (1999).

\bibitem{stegner}  A. Stegner and J.E. Wesfreid, Phys. Rev. E {\bf 60},
R3487 (1999).

\bibitem{cook}  H.E.~Cook, {\it Acta Met}.~{\bf 18}, 297 (1970).

\bibitem{bray}  I.M.~Lifshitz and V.V.~Slyozov, J.~Phys.~Chem.~Solids {\bf 19%
}, 35 (1961). A.J.~Bray, Adv.~Phys.~{\bf 43}, 357 (1994).

\bibitem{copol}  S.~Joly, A.~Raquois, F.~Paris, B.~Hamdoun, L.~Auvray,
D.~Ausserre and Y.~Gallot, {\it Phys.~Rev.~Lett.}~{\bf 77}, 4394 (1996).

\bibitem{gunton}  J.D.~Gunton, M.~San Miguel and P.S.~Sahni, in {\it Phase
Transition and Critical Phenomena}, edited by C.~Domb and J.L.~Lebowitz
(Academic, London, 1983), Vol.~8, p.~267.

\bibitem{baron}  J.S.~Langer, M.~Bar-on and H.D.~Miller, {\it Phys.~Rev.~A}, 
{\bf 11}, 1417 (1975).

\bibitem{saxena}  A.~Saxena and A.R.~Bishop, {\it Phys.~Rev.~A} {\bf 44},
R2251 (1991).

\bibitem{novik}  A.~Novik-Cohen and L.A.~Segel, {\it Physica D} {\bf 10},
277 (1984).

\bibitem{abra}  M.\thinspace Abramowitz and I.\thinspace Stegun, Handbook of
Mathematical Functions (Dover, New York,\thinspace 1965).

\bibitem{arscott}  F.M. Arscott, {\it Periodic Differential Equations}
(Pergamon, Oxford, 1981).

\bibitem{annexe}  $\frac{\partial }{\partial \phi }$ $\Phi (x,k,\phi =0)$
corresponds to a local maximum of the free energy averaged over one period $%
{\cal F}(\phi )=\int F(\Phi (x,k,\phi )){\rm d}x$. The (G-L) or stationary
(C-H) equation, i.e. the first functional derivative $\frac{\delta F}{\delta
\Phi }=0$, admits $\Phi (x,k,\phi =0)=\Psi ^{\ast }(x)$ as solution. But
Lam\'{e} equation, obtained when linearizing (G-L) equation\ around $\Psi
^{\ast }(x)$, is related with the second functional derivative of $F$;
therefore one doesn't expect $\frac{\partial }{\partial \phi }$ $\Phi
(x,k,\phi =0)$ to be an exact solution or eigenfunction of ${\cal L}$.

\bibitem{goldstein}  P.~Coullet, R.E. Goldstein and J.H.~Gunaratne,~Phys.
Rev. Lett~J.~63, 1954 (1989).
\end{references}
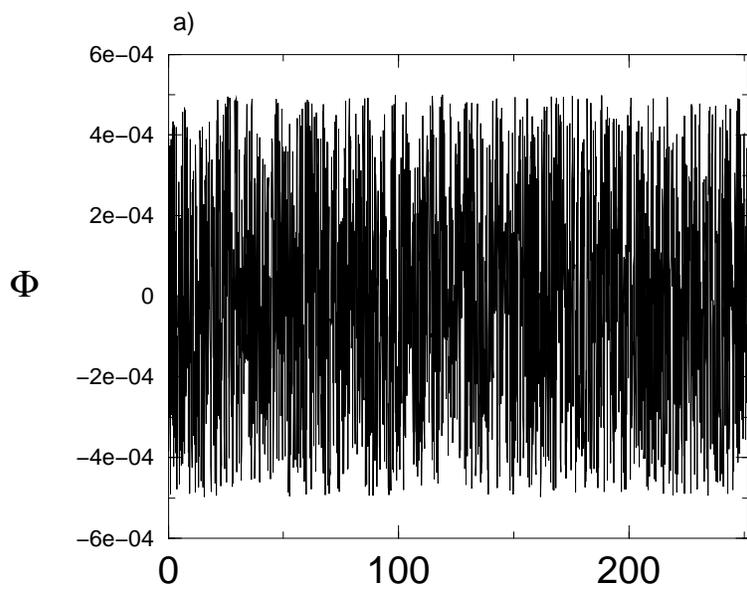
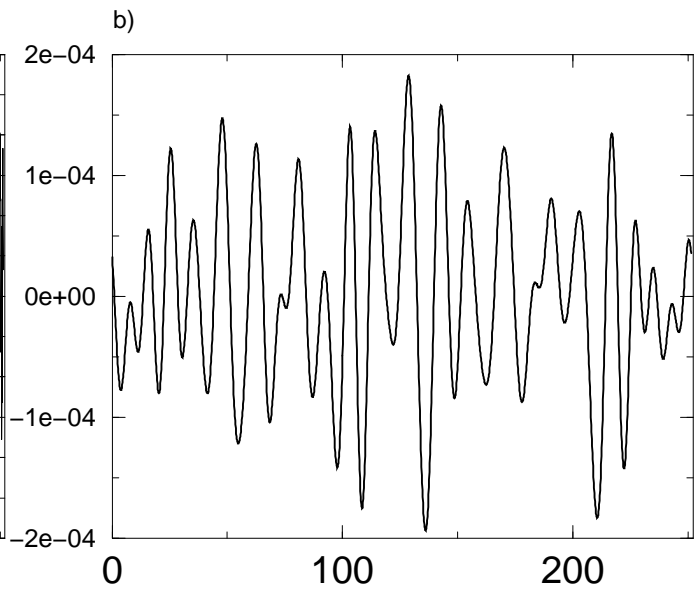
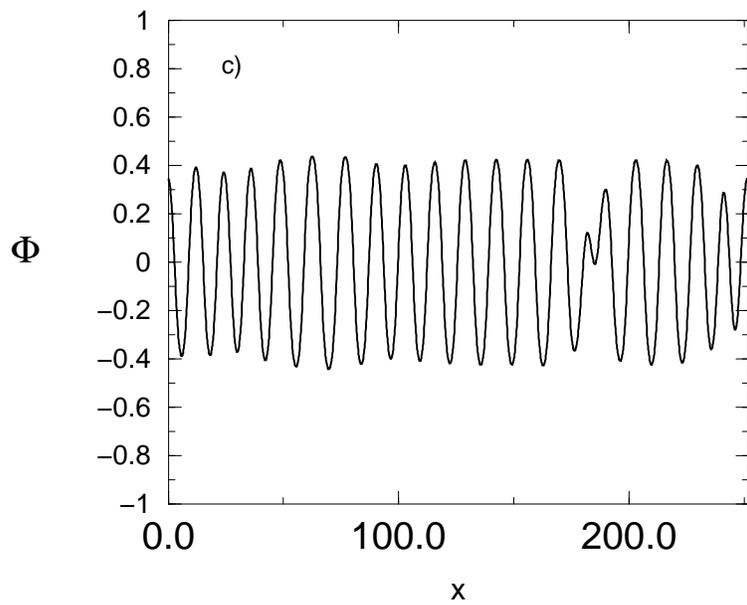
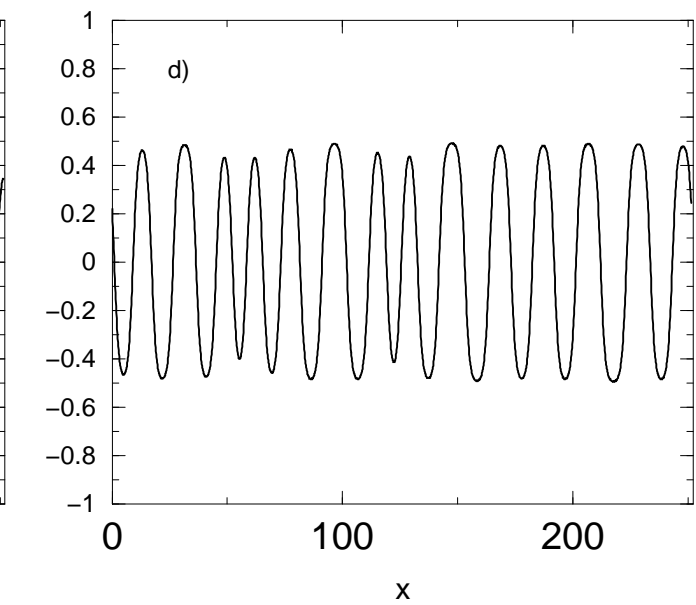
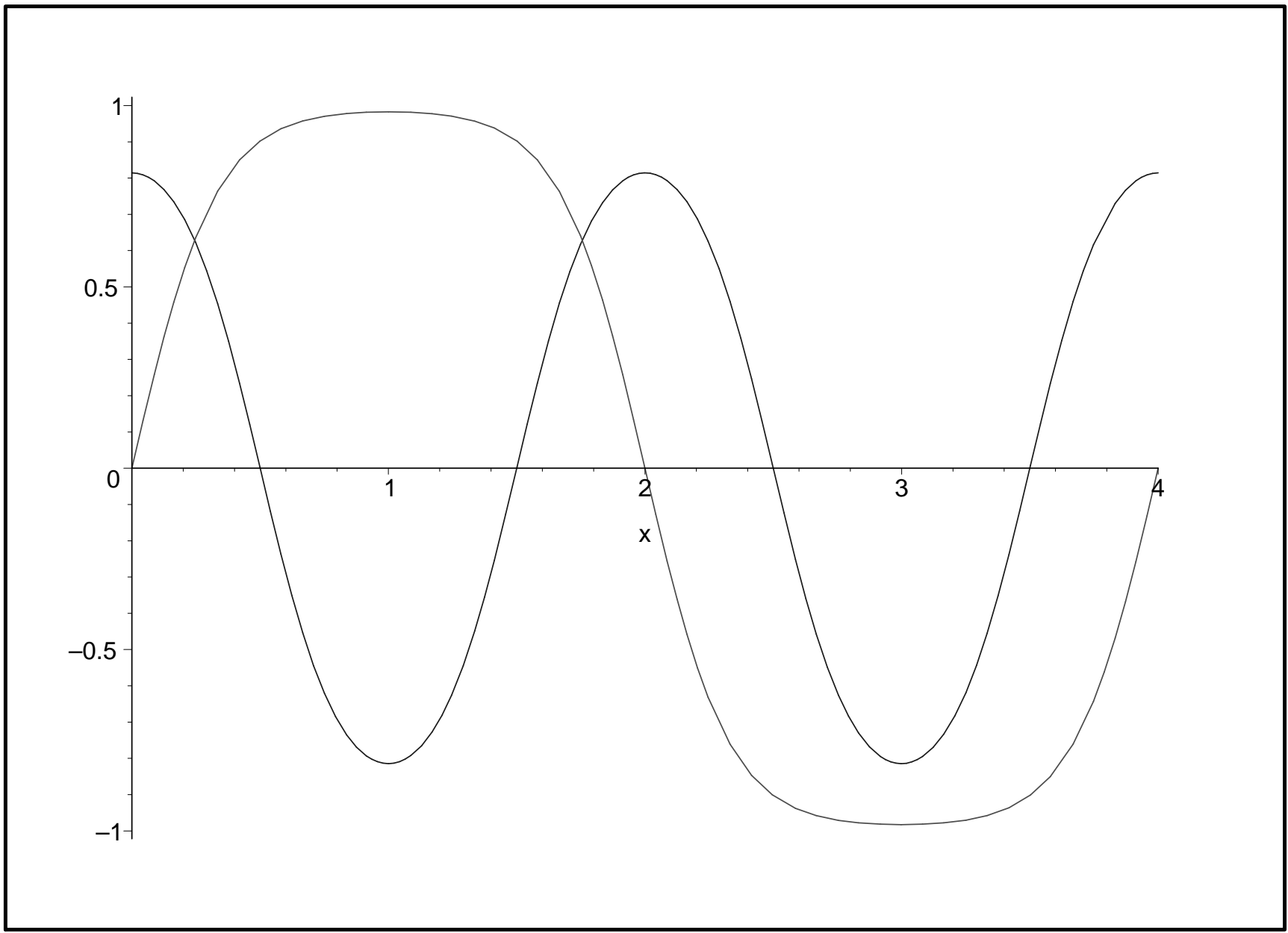
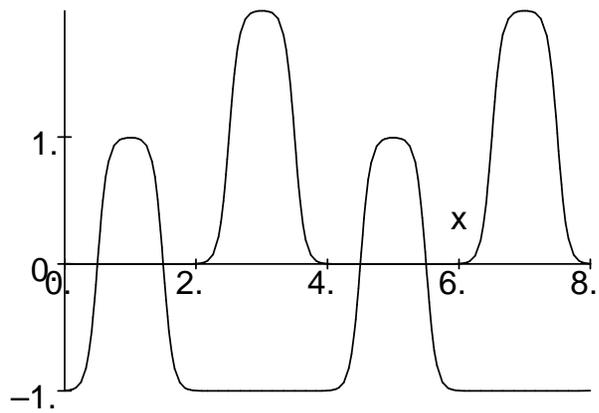
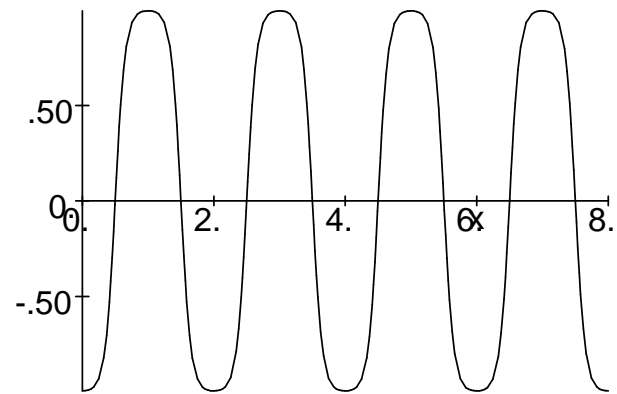
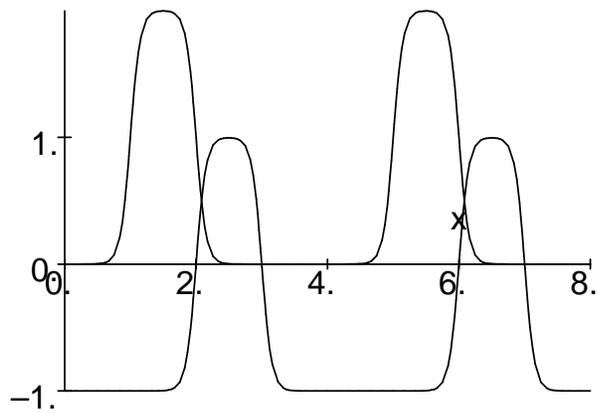
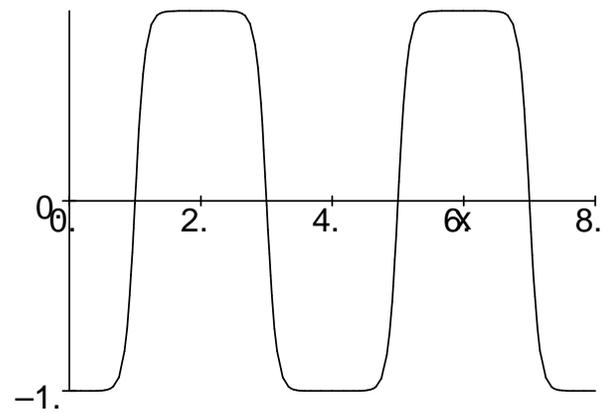
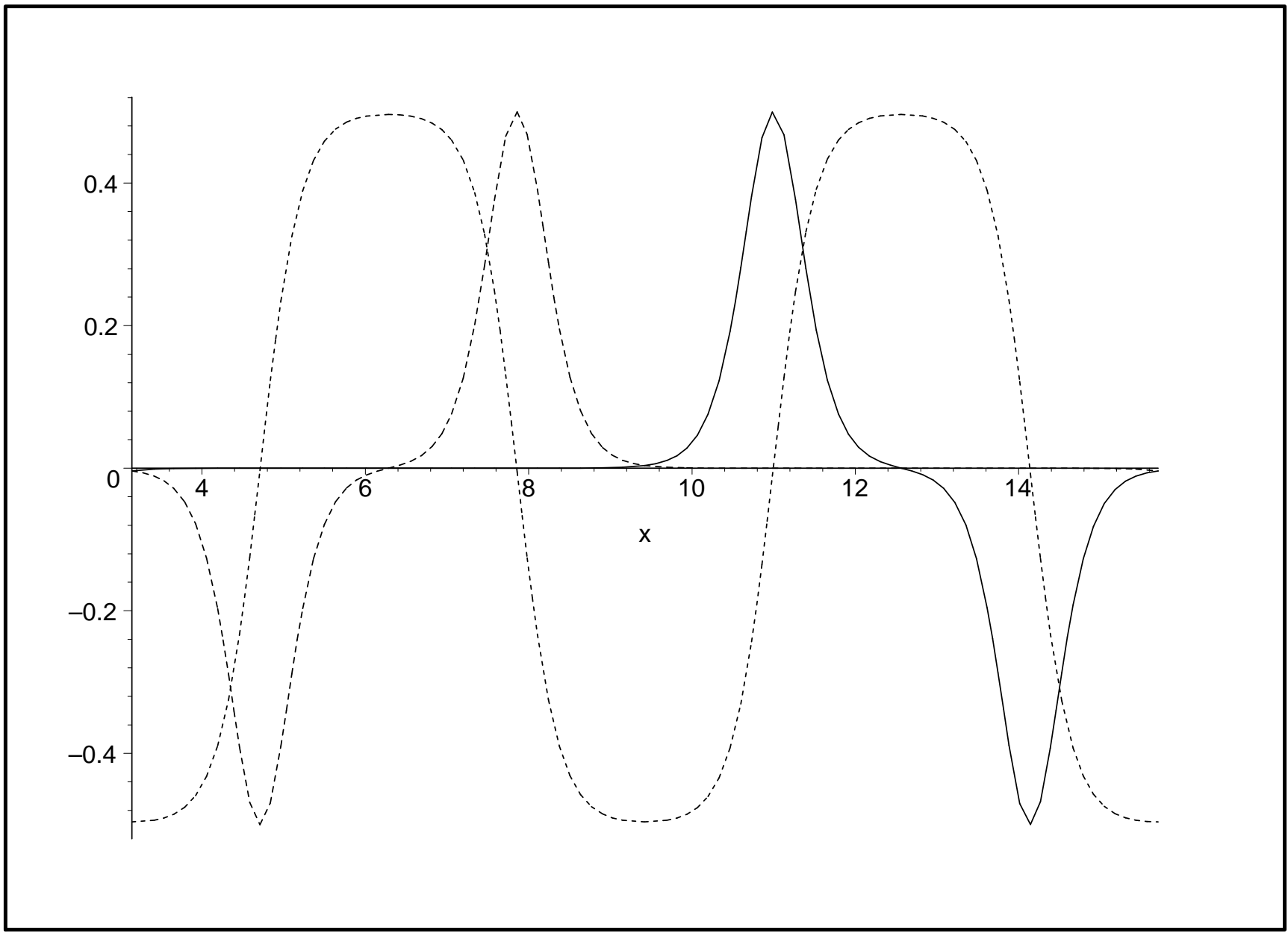
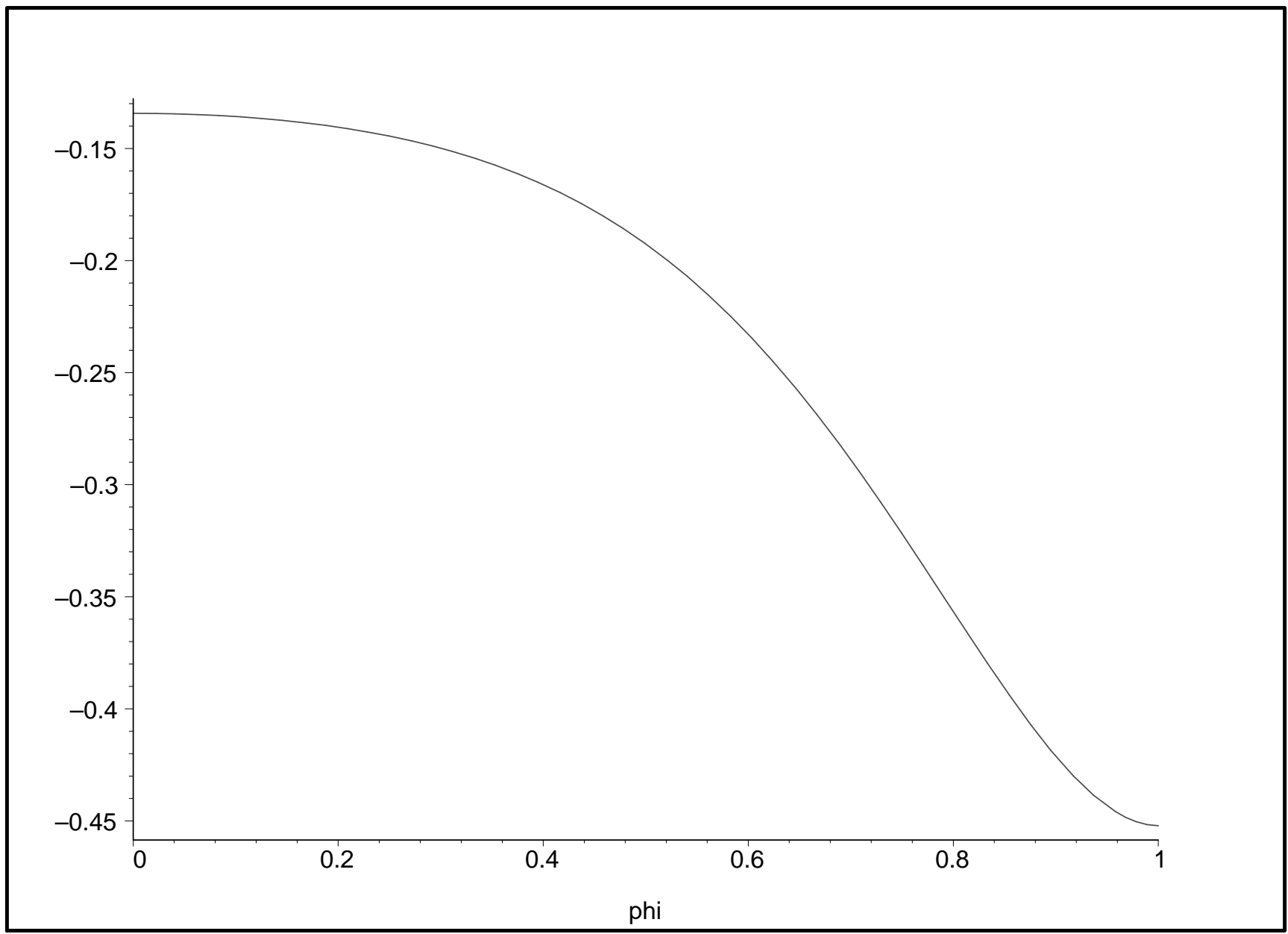
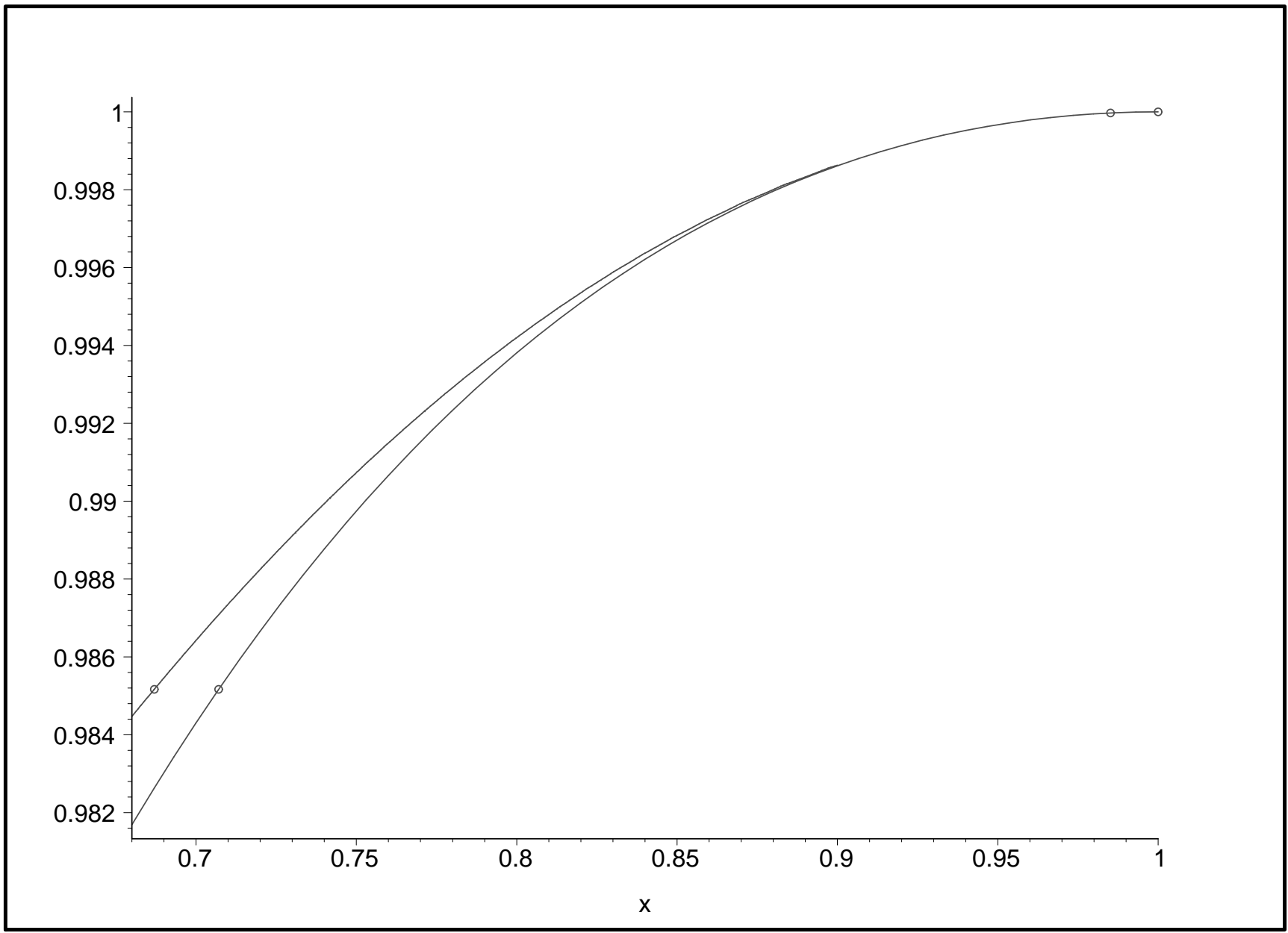
\end{document}